\documentclass[twocolumn,prl,showpacs,superscriptaddress,amsmath,citeautoscript]{revtex4}
\usepackage{graphicx}
\usepackage{bm}
\usepackage{amssymb}

\begin{document}

\title{Negative nonlocal resistance in mesoscopic gold Hall bars:\\ Absence of giant spin Hall effect}

\author{G.\ Mihajlovi\'{c}}
\email{mihajlovic@anl.gov}
\affiliation{
Materials Science Division, Argonne National Laboratory, Argonne, IL 60439\\}
\author{J.\ E.\ Pearson}
\affiliation{
Materials Science Division, Argonne National Laboratory, Argonne, IL 60439\\}
\author{M.\ A.\ Garcia}
\affiliation{
Dpto. F\'{\i}sica de Materiales, Universidad Complutense de Madrid, 28040 Madrid, Spain\\}
\author{S.\ D.\ Bader}
\affiliation{
Materials Science Division, Argonne National Laboratory, Argonne, IL 60439\\}
\affiliation{
Center for Nanoscale Materials, Argonne National Laboratory, Argonne, IL 60439\\}
\author{A.\ Hoffmann}
\affiliation{
Materials Science Division, Argonne National Laboratory, Argonne, IL 60439\\}
\affiliation{
Center for Nanoscale Materials, Argonne National Laboratory, Argonne, IL 60439\\}

\date{\today}


\begin{abstract}
We report the observation of negative nonlocal resistances in multiterminal mesoscopic gold Hall bar structures whose characteristic dimensions are larger than the electron mean-free path. Our results can  only be partially explained by a classical diffusive model of the nonlocal transport, and are not consistent with a recently proposed model based on spin Hall effects. Instead, our analysis suggests that a quasiballistic transport mechanism is responsible for the observed negative nonlocal resistance. Based on the sensitivity of our measurements and the spin Hall effect model, we find an upper limit for the spin Hall angle in gold of 0.022 at 4.5 K.
\end{abstract}

\maketitle

The term nonlocal resistance, $R_{nl}$, refers to the generation of a voltage in regions of a multiterminal structure that are outside of the nominal current path. In high mobility, two-dimensional semiconductor heterostructures, $R_{nl}$  and magnetoresistance (MR) measurements at low temperatures provided valuable insights into understanding electron transport in the ballistic \cite{Takagaki-Shepard,Hirayama} and quantum Hall regimes \cite{McEuen}. Such measurements were  also applied to study universal conductance fluctuations in diffusive semiconductors and metals in the phase-coherent regime \cite{Haucke}, and the flow of vortices in  superconducting channels \cite{Grigorieva}.  In hybrid ferromagnet/non-magnet lateral spin valve structures, nonlocal MR measurements have been utilized to study coherent spin transport phenomena, such as spin diffusion \cite{Ji,Jedema2003-Johnson}, spin precession \cite{Jedema-Lou}, and spin Hall effects (SHEs) \cite{Valenzuela-Kimura,Seki}.  Theoretically it has also been suggested that SHEs should give rise to experimentally observable $R_{nl}$ in purely paramagnetic structures when the magnitude of the spin Hall angle $\gamma$, defined as the ratio of spin Hall and charge conductivities, is sufficiently large \cite{Hankiewicz,Abanin}.  However, in this Letter, we present experimental results that show no signatures of such SHE induced $R_{nl}$ in mesoscopic structures fabricated from gold, which is inconsistent with the recently reported giant $\gamma$ in this material. We also report the surprising observation of a negative $R_{nl}$ and show this to be due to a quasiballistic charge transport mechanism not related to SHEs.

\begin{figure}
\includegraphics[scale=0.4, bb=0 9 595 425]{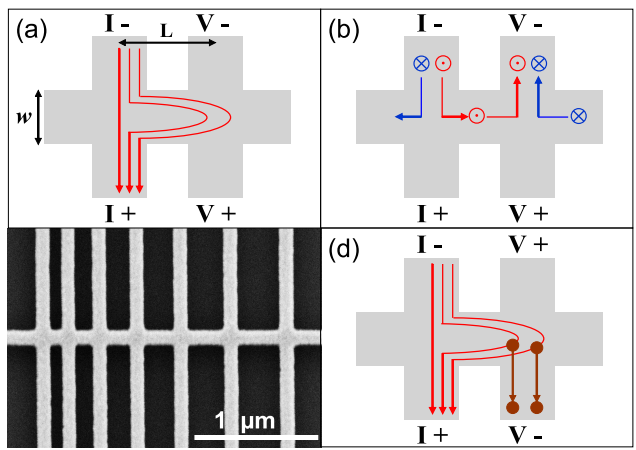}
\caption{(Color online) Schematic depiction of the physical mechanisms giving rise to (a) positive  $R_{nl}^c$, (b) positive $R_{nl}^{SH}$, and (d) negative $R_{nl}$ due to quasibalistic transport as described in the text. Arrows indicate the direction of electron flow. (c) SEM image of the central region of the gold Hall bar with seven vertical wires bridged by a horizontal one.}
\label{fig:schematics}
\end{figure}

Consider a Hall bar structure fabricated from a nonmagnetic normal metal, with two parallel vertical wires of width $w$ separated by a distance $L$ and  bridged by a horizontal wire of identical width (see Fig. 1), and at temperatures high enough that quantum effects can be neglected. When a current runs through one vertical wire and the voltage is measured across another, a non-zero $R_{nl}$ appears in the diffusive transport regime when the electron mean-free path $l_e \ll w$, because the current density, which spreads into the bridging wire, has a nonzero magnitude in the region between the voltage probes [see Fig.\ \ref{fig:schematics}(a)]. The magnitude of this classical $R_{nl}$, $R_{nl}^c$, decays exponentially with the distance between the wires at a rate set by the device geometry. Indeed, it follows from the van der Pauw theorem \cite{van der Pauw} that for $L \geq w$
\begin{equation}
  R_{nl}^c = R_{sq}\exp(-\frac{\pi L}{w}),
  \label{eq:diffusive}
\end{equation}
where $R_{sq} = \rho/t$ is the sheet resistance of the wire having resistivity $\rho$ and thickness $t$. $R_{nl}^c$ is positive, meaning the nonlocal voltage  has the same polarity as the one along the direction of current flow in the adjacent wire.

Recently, however, an additional transport mechanism, related to SHEs, has been predicted to give rise to nonzero $R_{nl}$ in a metallic Hall bar structure in the diffusive transport regime \cite{Abanin}. This mechanism is depicted schematically in Fig.\ 1(b); an electrical current flowing through the left vertical wire generates a perpendicular spin current in the bridging wire due to the direct SHE. $R_{nl}$ appears because the electrons carrying the spin current scatter preferentially in the same direction (inverse SHE) thus creating a charge accumulation, \emph{i}.\emph{e}. voltage across the right vertical wire. Abanin {\it et al.}\  calculated \cite{Abanin} that for $l_e \ll\ w \ll l_s$, where $l_s$ is the electron spin diffusion length, this $R_{nl}$ induced by SHEs, $R_{nl}^{SH}$, can be expressed as
\begin{equation}
  R_{nl}^{SH} = \frac{1}{2}\gamma^2 R_{sq} \frac{w}{l_s}\exp(-\frac{L}{l_s}).
  \label{eq:two}
\end{equation}
$R_{nl}^{SH}$ should also be positive (since the potential difference that builds up opposes the electron flow)  and, for a given $L$, may significantly contribute to $R_{nl}$, if $l_s$  and $\gamma$ are sufficiently large. It also follows from Eq. (2) that by analyzing $R_{nl}^{SH}$ as a function of $L$, one can simultaneously determine $l_s$  and $\gamma$. Compared to the experiments in which ferromagnets were used to generate or detect spin currents \cite{Valenzuela-Kimura,Seki}, this scheme offers an advantage of avoiding complications related to spin injection or detection efficiency of the ferromagnets, which must be known in order to determine $\gamma$.

Motivated by these predictions, we fabricated mesoscopic gold Hall bar structures with variable distance $L$ between adjacent vertical wires and measured $R_{nl}$ as a function of $L$ and temperature $T$. Gold was chosen due to its strong spin-orbit coupling, expected to give rise to a large $\gamma$ value, and at the same time a long enough $l_s$ value in order to provide a significant value of $R_{nl}^{SH}$ compared to $R_{nl}^{c}$. Indeed, it has been recently reported that a giant SHE exists in gold, with $\gamma$ = 0.113 at room temperature \cite{Seki}. Furthermore, values of $l_s$ of up to 168 nm at 10 K have been reported \cite{Ku}. Surprisingly, we observed that, in addition to  $R_{nl}^c$, the measured $R_{nl}$ contains a negative contribution  that  decays with $L$ exponentially, as predicted by a classical model, but whose magnitude is also proportional to the fraction of electrons that can travel ballistically over  the width of the bridging wire $w$, $i.e.$, $\propto \exp(-w/l_e)$. A similar mechanism had been previously observed to give rise to a negative value of $R_{nl}$ in structures fabricated from high mobility, two-dimensional electron systems in semiconductor heterostructures, but was typically neglected in nonlocal transport measurements on metallic nanostructures. In addition, based on Eq. (2) and the sensitivity of our measurements, we deduce $\gamma \leq 0.022$, which conflicts with the recent observation of a giant SHE in gold.

The Hall bar structures were fabricated on a SiN/Si substrate by e-beam lithography, e-beam evaporation and lift-off. A scanning electron microscopy (SEM) image of the central region is shown in Fig.~1(c). The width and the thickness of the wires were $w = (110 \pm 4)$ nm and $t = (60 \pm 2)$ nm, as determined by SEM and atomic force microscopy analysis respectively. The distance $L$ between the adjacent vertical wires was varied from 200 to 450 nm in 50 nm steps. All resistance measurements were performed by running an alternating {\it dc} current $I = \pm~0.5$ mA and measuring a {\it dc} voltage with a nanovoltmeter. The resistivity of the gold wires was 2.07~$\mu\Omega$cm and 3.89~$\mu\Omega$cm at 4.5 and 295~K, respectively. The corresponding values of $l_e$, calculated according to the Drude formula $l_e = (\hbar/e^2\rho)(3\pi^2/n^2)^{1/3}$ were 40.5 and 21.6~nm, respectively, when using an electron density for gold of $n = 5.9 \times 10^{28}$~m$^{-3}$.

\begin{figure}
\includegraphics[scale=0.39, bb=40 30 605 495]{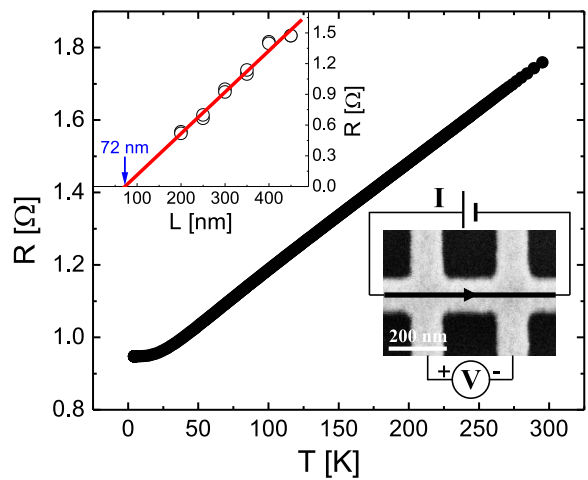}
\caption{(Color online) Resistance of the bridging wire ($L = 300$~nm) as a function of $T$. Lower inset: SEM image of the segment of the wire whose resistance is plotted, adapted to show the actual measurement configuration. Upper inset: Resistance of each segment of the bridging wire between adjacent vertical wires (open circles) at 4.5 K as a function of  $L$. The red line is a linear fit to the data.}
\label{fig:resistance}
\end{figure}

Figure 2 shows the local resistance, $R = V/I$, measured between the two vertical wires, separated by 300 nm along the current path (see right inset), from 4.5 to 295 K. In addition to the $T$ dependence of $R$, we also measured the resistance for each segment of the bridging wire between the adjacent vertical wires at 4.5 K. The plot of $R$ as a function of $L$ is shown in the inset of Fig.~2. As expected, $R$ increases linearly with $L$. However, the linear fit crosses the $L$-axis at $L_0 = (72 \pm 17)$~nm, which indicates that the effective distance between the vertical wires, $L_{eff} = L - L_0$, is shorter than $L$. This is most likely caused by spreading of the current density into the voltage leads, due to their finite width.

\begin{figure}
\includegraphics[scale=0.39, bb=40 30 605 495]{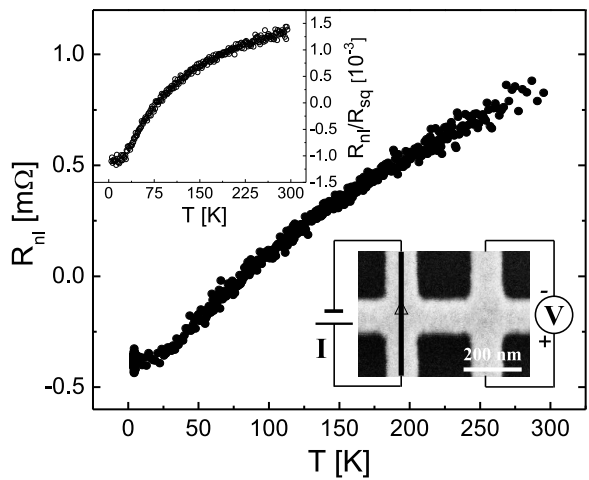}
\caption{Temperature dependence of $R_{nl}$ measured for wires separated by a distance $L = 300$~nm. Lower inset: SEM image of the segment of the structure where $R_{nl}$ was measured, adapted to show the actual measurement configuration.Upper inset: $R_{nl}/R_{sq}$ as a function of temperature.}
\label{fig:nonlocal}
\end{figure}

In contrast to $R(T)$, the $T$ dependence of the nonlocal resistance is unexpected. Figure 3 shows $R_{nl}$ $vs.$ $T$ data obtained for $L = 300$~nm. For other combinations of adjacent vertical wires the data show the same qualitative behavior, but different overall magnitude. At room temperature $R_{nl}$ is positive, as expected based on the classical $R_{nl}$ mechanism. When $T$ is lowered $R_{nl}$ decreases, but this decrease is not proportional to the decrease of $R_{sq}$, as expected based on Eq.~(1). This can be seen in the inset of Fig.~3 where we plot $R_{nl}/R_{sq}$ $vs.$ $T$. Based on Eq.~(1), $R_{nl}/R_{sq}$ should be a $T$ independent constant, determined solely by the  geometry of the structure.  However, we observed that $R_{nl}/R_{sq}$ is strongly $T$ dependent. Even more surprisingly, $R_{nl}$ changes sign with decreasing $T$, becoming negative around 82~K, as can be seen in the main plot of Fig.~3. The appearance of the negative $R_{nl}$ rules out that the SHE mechanism, as suggested in Ref.~\onlinecite{Abanin}, is responsible for the observed $T$ dependence of $R_{nl}/R_{sq}$.

What transport mechanism could be responsible for the appearance of a $T$ dependent negative  $R_{nl}$? Previous numerical studies  \cite{Takagaki94} suggest that a negative  $R_{nl}$ can appear due to direct ballistic transmission of electrons into the voltage lead. This purely classical mechanism, however, is expected to be relevant only in structures fabricated from high mobility semiconductor heterostructures where $l_e$  exceeds  characteristic dimensions of the structure. Indeed, negative $R_{nl}$ has been observed in modulation doped GaAs Hall bar structures in the ballistic transport regime \cite{Hirayama}. In order to analyze whether ballistic electrons are responsible for negative $R_{nl}$ in our structures, we plot $R_{nl}/R_{sq}$ as a function of  $l_e$. The plots are shown on Fig. 4(a)-(f) for all distances $L$ between the adjacent vertical wires. We find that all the data can be fitted to the formula
\begin{equation}
  R_{nl}/R_{sq} = a[1-b\exp(-\frac{w}{l_e})],
  \label{eq:three}
\end{equation}
where $a$ and $b$ are dimensionless fitting parameters. Figure 5(a) shows the plot of $a$ values extracted from the fit as a function of $L$. The fitting curve on the graph  corresponds to $a = \exp[-\pi (L - L_0)/w]$ with $L_0 = (71.1 \pm 0.4)$~nm, in excellent agreement with $L_0$ obtained from the local resistance measurements. Therefore, in the completely diffusive limit, $l_e \rightarrow 0$, we recover the classically expected $R_{nl}^{c}$ with $L = L_{eff}$ [see Eq.~(1)]. In addition, it follows that the negative contribution to $R_{nl}$, since it is  also proportional to $\exp(-\pi L_{eff}/w)$, must originate from the spreading of the current density into the bridging wire. On the other hand, the term $\exp(-w/l_e)$ in Eq.~(3) is the fraction of  electrons that can travel ballistically over the distance $w$. Thus, we conclude that the negative $R_{nl}$ comes from electrons that reach the region between the voltage probes diffusively and then ballistically scatter into the lower voltage lead, generating a negative voltage. The proposed mechanism is depicted schematically in Fig. 1(d).

\begin{figure}
\includegraphics[scale=0.48, bb=0 0 495 475]{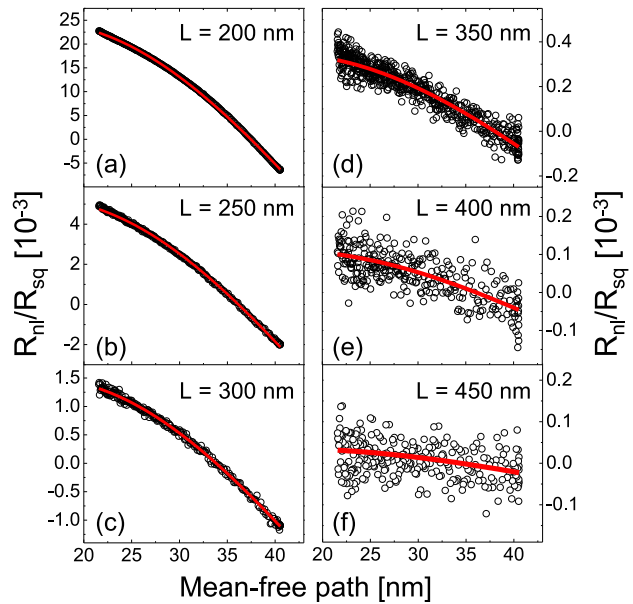}
\caption{(Color online) (a)-(f) $R_{nl}/R_{sq}$ as a function of electron mean-free path for different $L$ values. Corresponding fits to Eq.~(3) are shown as red lines.}
\label{fig:mfp}
\end{figure}

The values of $b$ extracted from fitting the $R_{nl}/R_{sq}$ $vs.$ $l_e$ curves are plotted in Fig. 5(b) for different distances $L$. $b$ varies periodically with $L$, which suggests that it is somewhat sensitive  to the rebound electron trajectories \cite{Takagaki94}. The oscillatory behavior of the negative $R_{nl}$ further supports the explanation that ballistic electrons are responsible for its appearance.

Finally we subtract $R_{nl}^c$ from the measured $R_{nl}$ and plot the data points as a function of $L_{eff}$ for two different temperatures corresponding to the upper (295~K) and lower (4.5~K) limit of our measurement range. The data were fitted to an exponential function  $\Delta R_{nl} = \alpha \exp(-L_{eff}/\lambda)$ with fixed $\alpha = - R_{sq}\langle b\rangle\exp(-w/l_e)$, where  $\langle b\rangle$ = 21.5 is the average value of $b$, and  $\lambda$ as a fitting parameter. Both data plots and the corresponding fitting curves are shown in Fig. 5(c). We obtain $\lambda = (33.0 \pm 0.9)$~nm and $\lambda = (34.0 \pm 0.2)$~nm at 295 and 4.5 K, respectively. Therefore, there is no difference in the dependence of the negative  $R_{nl}$ on $L_{eff}$ at different temperatures. One can also see that $\lambda = w/\pi$, which further confirms that the classical diffusive transport mechanism is partially responsible for the appearance of our negative $R_{nl}$ values.

\begin{figure}
\includegraphics[scale=0.42, bb=10 1 570 340]{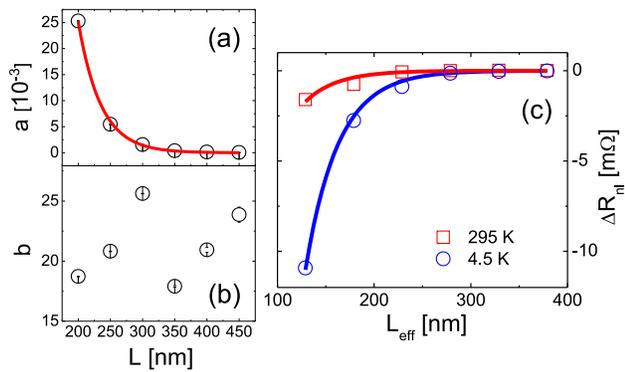}
\caption{(Color online) (a) Parameter $a$ $vs.$ $L$ (open circles) and the best exponential fit (red line) as explained in the text; (b) Parameter $b$ $vs.$ $L$; (c) Negative $R_{nl}$ $vs.$ $L_{eff}$ at 4.5 and 295~K and the corresponding fitting curves explained in the text.}
\label{fig:s2l2}
\end{figure}

Thus, our experiments do not show the additional positive contribution to $R_{nl}^{c}$ expected from SHEs.  Mainly, such a positive contribution should manifest itself by a different dependence of $R_{nl}$ on $L_{eff}$ at different temperatures, due to the $T$ dependence of $l_s$ \cite{Otani}. Using the resolution of our measurements, which is 20~$\mu\Omega$, and Eq.~(2) with $L = L_{eff}$ = 128.9 nm (which corresponds to shortest separation between the vertical wires in our structure) we can deduce an upper limit for  $\gamma$  of 0.022 at 4.5 K assuming  $l_s = 65$~nm \cite{Ji}, and 0.027 at 295 K assuming  $l_s = 36$~nm. Note that the values of $\gamma$ in both cases would be even lower if  $l_s$ were longer. Therefore, based on our experiments, the values of $\gamma$ are at least an order of magnitude lower than the ones reported in Ref.~\onlinecite{Seki}. A possible reason for this discrepancy could be the recent suggestion that Fe impurities and their concomitant Kondo effect could be responsible for a giant $\gamma$ in gold \cite{Nagaosa}.  To this end, we note that we have not observed an upturn in the resistivity of our structures down to the lowest $T$, which would be a signature of such a Kondo effect (see Fig. 2). Another possible reason why a giant SHE was inferred in Ref.~\onlinecite{Seki} could be a strong sensitivity of the nonlocal charge signal to local stray magnetic fields from ferromagnetic components of the structure \cite{Monzon-van Wees}.  It is worthwhile to note that we observe significant $R_{nl}$ signals, independent of spin transport, in structures with comparable geometries and dimensions as the ones studied in Ref.~\onlinecite{Seki}, but their dependence on local magnetic fields is not clear at this point.  In general, these effects, that are based only on charge transport, have been neglected in most experiments aimed at nonlocal detection of spin currents.

In conclusion, we report a negative value of  $R_{nl}$ in mesoscopic Hall bar structures fabricated from a nonmagnetic metal, $i. e.$ gold. Our analysis shows that negative $R_{nl}$ value arises from the effect of ballistic electrons on nonlocal transport, despite the fact that the electronic mean-free paths are smaller than the dimensions of the structure. In addition, our results do not support the recently reported giant spin Hall effect in gold.

We thank Dimitrie Culcer, Roland  Winkler, Oleksandr Mosendz, and  Sadamichi Maekawa for useful discussions and comments.  This work was supported by
DOE BES under Contract No. DE-AC02-06CH11357.


\end{document}